%% file: main.tex
\documentclass[preprint]{article}

\usepackage{neurips_2025}

\usepackage[utf8]{inputenc} % allow utf-8 input
\usepackage[T1]{fontenc}    % use 8-bit T1 fonts
\usepackage{hyperref}       % hyperlinks
\usepackage{url}            % simple URL typesetting
\usepackage{booktabs}       % professional-quality tables
\usepackage{amsfonts}       % blackboard math symbols
\usepackage{nicefrac}       % compact symbols for 1/2, etc.
\usepackage{microtype}      % microtypography
\usepackage{xcolor}         % colors
\input{contents/usepackages}

\title{\Large On the Practice of Scaling Search Conversion Rate Prediction}

\author{
    James Pak \quad Jyun-Yu Jiang \quad Fan Zhang \quad Sen Wang \quad Taekmin Kim \quad Henry Tsai\\
    \bf 
    Vijay Rajaram \quad Juexin Lin \quad Mohitdeep Singh \quad Alessandro Magnani \quad Johnny Chen \\
    \bf 
    Qian Zhao \quad Rao Fu \quad Zhirong Liang \quad Jordan Gilliland \quad Winter Jiao \\
    Coupang\\
    \texttt{\{mipak, jyjiang, fazhang1, sewang1, takim856, mitsai, vijay, julin9}\\
    \texttt{mosingh, almagnan, jichen9, qizhao, rao, zhliang, cjg, dojiao\}@coupang.com}\\
}

\if 0
\author{%
    James Pak\\
    Coupang\\
    \texttt{mipak@coupang.com}\\
    \And
    Jyun-Yu Jiang\\
    Coupang\\
    \texttt{jyjiang@coupang.com}\\
    \And
    Fan Zhang\\
    Coupang\\
    \texttt{fazhang1@coupang.com}\\
    \And
    Sen Wang\\
    Coupang\\
    \texttt{sewang1@coupang.com}\\
    \And
    Taekmin Kim\\
    Coupang\\
    \texttt{takim856@coupang.com}\\
    \And
    Henry Tsai\\
    Coupang\\
    \texttt{mitsai@coupang.com}\\
    \And
    Vijay Rajaram\\
    Coupang\\
    \texttt{vijay@coupang.com}\\
    \And
    Juexin Lin\\
    Coupang\\
    \texttt{julin9@coupang.com}\\
    \And
    Mohitdeep Singh\\
    Coupang\\
    \texttt{mosingh@coupang.com}\\
    \And
    Alessandro Magnani\\
    Coupang\\
    \texttt{almagnan@coupang.com}\\
    \And
    Johnny Chen\\
    Coupang\\
    \texttt{jichen9@coupang.com}\\
    \And
    Qian Zhao\\
    Coupang\\
    \texttt{qizhao@coupang.com}\\
    \And
    Rao Fu\\
    Coupang\\
    \texttt{rao@coupang.com}\\
    \And
    Zhirong Liang\\
    Coupang\\
    \texttt{zhliang@coupang.com}\\
    \And
    Jordan Gilliland\\
    Coupang\\
    \texttt{cjg@coupang.com}\\
    \And
    Winter Jiao\\
    Coupang\\
    \texttt{dojiao@coupang.com}\\
}
\fi

\begin{document}

\maketitle

\begin{abstract}
Scaling a Search Conversion Rate (CVR) prediction model, especially in high-traffic environments, presents a challenge: superior model quality needs to be balanced with strict constraints on training cost and serving latency. This paper details an effective approach for scaling modern search CVR prediction models. We begin with an empirical study to understand the scaling performance of search CVR models, analyzing how quality improves as we scale three key factors of model backbone computation, the size of embedding parameters, and the volume of training data.

We use a large-scale production dataset, comprising over a year of customer interaction logs from a high-traffic e-commerce platform, to evaluate the scalability of several state-of-the-art architectures and their ensembles. Our key findings are: (1) selecting the right backbone and scaling factors is crucial; (2) the impact of scaling backbone, embedding, and data is largely independent and additive, which has implications for more efficient scaling exploration; (3) a streamlined warmstart strategy can accelerate training iterations while simplifying new updates; (4) inference optimization strategies such as decoupled graph execution and dynamic batching can enable low-latency GPU serving even for high-capacity models.
Compared to a baseline of a pre-scaling production model, we ultimately deployed a model trained on 2.5x larger training data with 8x more inference compute while having minimal latency impact.
Online A/B tests also demonstrate that our launches achieved a combined +2.6\% gain in a key metric of search conversion rate.
\end{abstract}

\section{Introduction}
\label{section:introduction}
\input{contents/section1_introduction}

\input{contents/section2_prelim}

\section{Efficient Model Scaling}
\label{section:efficientmodeling}
\input{contents/section3_efficientmodeling}

\section{Experiments}
\label{section:exp}

\input{contents/section4_exp}

\input{contents/section5_relatedwork}

\input{contents/section6_conclusion}

\bibliographystyle{plain}
\bibliography{reference}

\end{document}

%% file: contents/usepackages.tex
\usepackage{algorithmic}
\usepackage[ruled,vlined, linesnumbered]{algorithm2e}
\usepackage{amsmath}
\usepackage{amssymb}
\usepackage{amsthm}

\usepackage{bm}
\usepackage{bbm}
\usepackage{enumitem}
\usepackage{subcaption}
\usepackage{textcomp}
\usepackage{multirow}
\usepackage{hyperref}
\usepackage{makecell}
\usepackage{tikz}
\usepackage{pgfplots}
\usepackage{pgfplotstable}
\usepgfplotslibrary{groupplots}
\pgfplotsset{compat=1.18}
\usepackage{seqsplit}

\newcommand{\mysection}[1]{\smallskip \noindent {\bf #1.}}

 \usepackage{amsthm}
\theoremstyle{definition}

\usepackage{lipsum}

\if 0

\fi

%% file: contents/section1_introduction.tex
Conversion Rate (CVR) prediction is central to search and recommendation systems, in various domains including e-commerce~\cite{cheng2016wide}.
While deep learning has already delivered significant gains in model quality ~\cite{zhuang2025practice}, its true potential lies in its ability to scale, enabling more complex architectures and better generalization over massive datasets. However, effectively realizing these scaling benefits remains a challenge, particularly in the search domain, where it is less well-studied.

%Unlike large language models (LLMs) whose scaling laws often show predictable improvements with scale~\cite{kaplan2020scaling}, CVR models are more difficult to tame. 
Scaling CVR models is challenging as many studies demonstrate that incremental returns usually diminish rapidly along scaling~\cite{yan2025scaling}.
First, the architecture of CVR prediction models is highly heterogeneous to model diverse inputs, so the scaling effects would be misaligned across different components like diverse scaling factors and embeddings. 
The embedded-centric architecture also introduces the majority of parameters concentrated in large embedded tables used to represent categorical features~\cite{naumov2019deep}.
Second, the complex model inputs comprise a mixture of categorical and continuous features from customer profiles, query context, and item attributes.
CVR models could have a hard time dealing with the balance between memorization and generalization.
The heterogeneity and amount of these features also complicate the preprocessing and generation of large volume of training data.
Lastly, online CVR systems operate under both strict inference latency and resource constraints, precluding the use of excessively slow or large architectures.

A promising approach for recommendation systems to achieve more predictable model scaling is to replace engineered engagement features with sequences of raw engagement actions~\cite{sun2019bert4rec, zhai2024actions, yan2025scaling}. Directly modeling the full history of raw engagement actions across all users presents a high degree of computational challenges. Moreover, search CVR presents different challenges compared to non-search recommendation problems. search contexts involve numerous point-in-time features and specific query intents that may not generalize well to sequential inputs for other sessions.
For example, the price and availability of a certain item are important features that can vary over time, or a current customer query could be completely irrelevant to previous searches.
Lastly, for both search and recommendation CVR models, the architectural backbones of many state-of-the-art frameworks are not inherently aligned with those of language models, which have shown predictable scaling~\cite{zhang2022dhen,zhuang2025practice}.

In this paper, our objective is to uncover a generalizable and robust scaling mechanism for search CVR models through systematic experimentation.
To this end, we comprehensively evaluate several leading state-of-the-art architectures and their ensemble variants.
Specifically, we decompose the models into three primary scaling dimensions: backbone, embeddings, and data, followed by an in-depth analysis of the scalability properties and effective scaling recipes for each dimension.
%In addition, we also propose to leverage the findings based on data scaling as a practical means to expedite model selection.
Finally, we demonstrate that inference optimizations can unlock the ability to serve the resulting scaled models with minimal cost or latency impact. We verify the effectiveness and real-world impact of the resulting models through large-scale online A/B experiments.
The main contributions of our work are summarized as follows:
\begin{itemize}[leftmargin=*]
%[leftmargin=*, topsep=\baselineskip, partopsep=0pt]
  \item We methodically investigate the scalability properties and practical impact of each scaling factor in state-of-the-art search CVR models. Our findings provide generalizable guidance for practitioners, helping them prioritize exploration efforts when scaling models in production settings.
  \item Leveraging the observed scaling properties of the data dimension, we propose a practical strategy to accelerate model selection: first conduct architecture search and hyperparameter optimization on smaller datasets, then transfer the selected configuration to the full-scale dataset for final training and deployment.
  \item We propose a streamlined warm-starting technique during model training that significantly reduces the time required for routine model development in production.
  \item To support the deployment of large-scale CVR models, we also present a practical framework for optimizing GPU efficiency in real-world serving environments.
  \item Experimental results on a large-scale customer dataset confirm the validity of our analysis and the effectiveness of the proposed scaling mechanisms. Furthermore, large-scale online A/B testing demonstrates tangible real-world improvements across multiple key success metrics, including a +2.6\% increase in search customer conversion rate.
\end{itemize}
\if 0
To scale up search CVR prediction models effectively, we need to address several challenges:
\begin{itemize}[leftmargin=*]
    \item Challenge 1 (C1): How does scaling up computation, parameter size, and data individually impact the quality and training efficiency of CVR prediction models?
    \item Challenge 2 (C2): Given the infrastructure limitations inherent in production serving environments, what is the most effective strategy to scale up a CVR prediction model?
\end{itemize}

This paper aims to provide a clear and practical path for scaling industrial CVR prediction models. The main contributions of our work are summarized as follows:
\begin{itemize}[leftmargin=*]
  \item We systematically investigate the scaling laws for search CVR prediction models with respect to computation cost, model size, and data size, providing a foundational understanding for the industry.
  \item We evaluate the computation scaling capabilities of several popular model architectures and share our practical lessons on how to successfully apply different models for CVR prediction tasks.
  \item We show that a horizontal ensembling approach is able to surpass the limitations of single-model scaling and propose efficient diversity-aware data sampling for data scaling.
  \item We share practical lessons on how to optimize inference speed and efficiency in production environments. 
\end{itemize}

\fi

\noindent \textbf{Disclaimer:}
Our framework is designed without any inclusion of customer personally identifiable information (PII). To ensure privacy, we utilize model features and training data that preclude the identification of any specific individual.

%% file: contents/section2_prelim.tex
\section{Preliminaries on Conversion Rate Prediction}
\label{section:prelim}

We first provide a preliminary overview of modern machine learning frameworks for CVR prediction.
In this work, we consider \texttt{purchases} as the target conversion behavior.
Given a query $q$ submitted by a customer $u$ and a set of exposed candidate items $D = \{d_i\}$, the goal of the CVR prediction task is to produce a ranking score $f(u, q, d_i)$ for each candidate item $d_i$ by modeling the probability that the customer $u$ would convert to a purchase of $d$:
$\mathbb{P}\left(\text{Purchase} \mid u, q, d_i\right)$.
Note that the problem statements in studies could slightly vary based on different business goals like advertising~\cite{zhuang2025practice} and recommendation~\cite{cheng2016wide}, but the general foundational mechanisms should be similar across business domains.

\begin{figure}[!t]
    \centering
    \includegraphics[width=.8\linewidth]{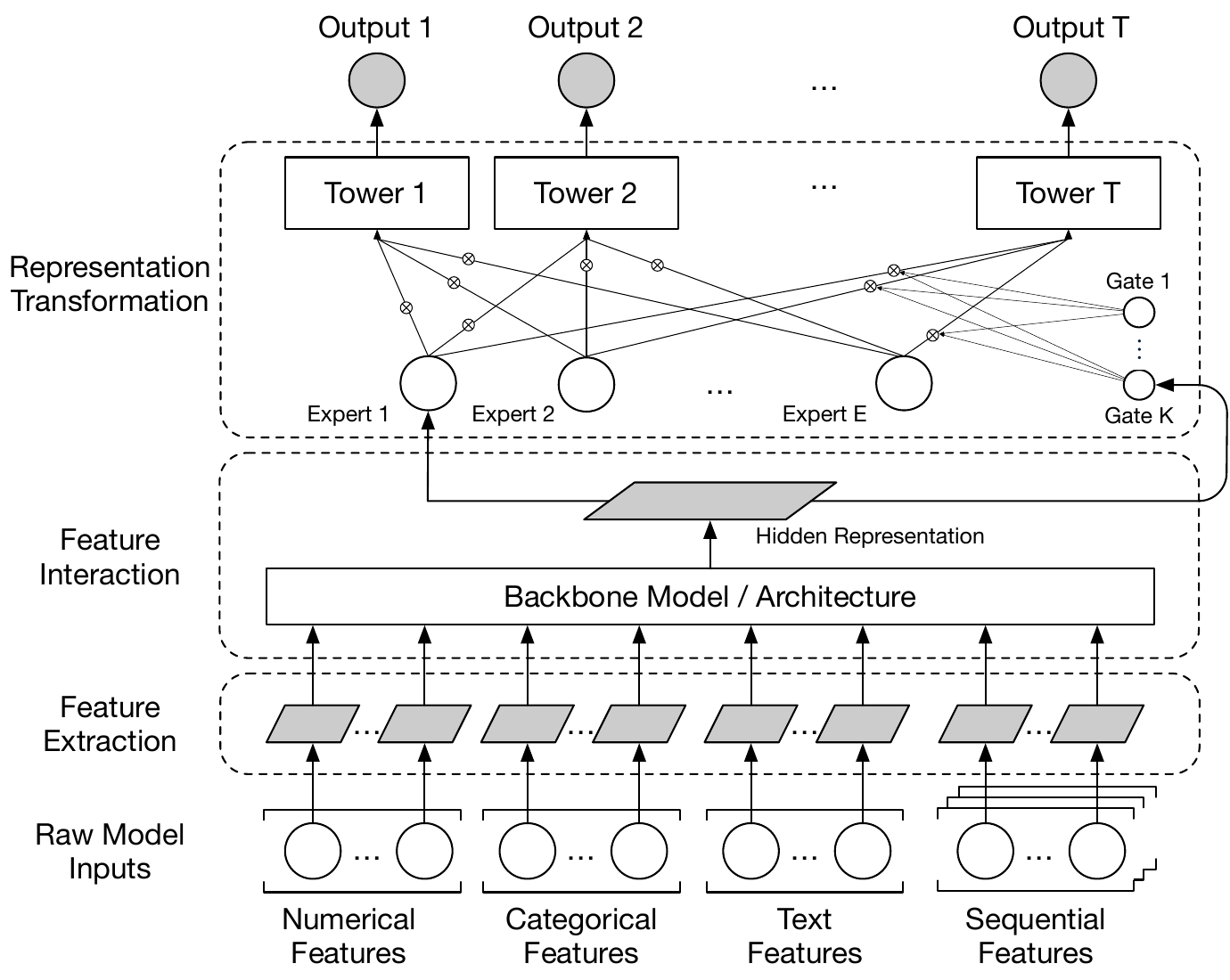}
    \caption{Overall CVR Model Framework.}
    \label{fig:overallframework}
    %\vspace{-16pt}
\end{figure}

\subsection{CVR Model Framework}
As shown in Figure~\ref{fig:overallframework}, most modern CVR prediction models can be generalized to three components, namely (1) feature extraction, (2) feature interaction, and (3) representation transformation~\cite{DLRM19}.

\mysection{Feature Extraction}
The raw inputs of a CVR model can be diverse and complex, including statistical features, text~\cite{ouyang2023contrastive}, contextual data~\cite{pan2022metacvr}, and many other attributes.
However, most of these inputs cannot be directly consumed by machine learning models since they could be non-standardized.
Accordingly, CVR models need to process and transform the raw inputs into dense vectors as machine-readable representations.

We categorize model inputs into numerical, categorical, text, and sequential features~\cite{cheng2016wide}, with the following conventional feature extraction methods:
\begin{itemize}[leftmargin=*]
    \item Each \textbf{numerical feature} is applied with normal standardization and \texttt{log1p} transformation~\cite{qin2021neural} as two normalized features.
    \item Each \textbf{categorical feature} is designated to an embedding layer to map discrete categories to continuous embedding vectors~\cite{cheng2016wide}.
    \item Each \textbf{text feature} is first tokenized into character n-grams~\cite{abeliuk2020que2search, gillick2019cld3} and then mapped into n-gram embeddings, followed by an average pooling layer to derive the text embedding.
    \item Each \textbf{sequential feature} is a temporal sequence of historical customer actions on items that is encoded into dense customer embeddings by sequence modeling for better personalization~\cite{zhuang2025practice}. Note that the customer embeddings do not utilize any personally identifiable information.
    
\end{itemize}
%To deal with the large cardinalities of categorical features and text n-grams, we also leverage the hashing trick~\cite{kang2021learning} to ameliorate the memory overhead associated with embedding tables. 
In this study, we focus on scaling multiple embedding features that are critical to search conversion rate prediction, such as query embedding and item embeddings.

\mysection{Feature Interaction}
The implicit semantics contained within the varied inputs are a critical part of CVR prediction. To capture the complex relationships among dense vectors from feature extraction, the feature interaction module functions as the cornerstone in modern CVR models.
Specifically, a backbone model or an ensemble of backbones, such as Deep \& Cross Networks ~\cite{wang2017deep, wang2021dcnv2} and Transformer~\cite{zhu2025rankmixer}, serves to map these dense embeddings into a rich, high-order hidden representation.
In this paper, we focus mainly on exploring several strategies to scale up the feature interaction component of CVR models as described in Section~\ref{section:efficientmodeling}.

\mysection{Representation Transformation}
To predict diverse customer behaviors, different forms of representations are needed in CVR prediction models. These are typically multi-task models in order to capture such behavior, combining conversion rate prediction with other correlated tasks such as click-through rate prediction (CTR). 
It is important to transform the shared output from the feature interaction module into a customized representation for each prediction task and capture task-specific information. In this paper, we choose the Multi-gate Mixture of Experts (MMoE) architecture~\cite{ma2018mmoe} for representation transformation. Note that this work does not include the study of this component, which is left to future work.

\if 0
To accommodate the prediction of multiple customer feedback, our model employs a multi-task learning framework. This requires transforming the shared output from the feature interaction module into specialized inputs customized for each prediction task. To achieve this, we adopt the Multi-gate Mixture of Experts (MMoE) architecture \cite{ma2018mmoe}, which effectively learns to route information and capture task-specific relationships from a shared representation. 

Please note that the scope of this work does not include alterations to this component.
\fi

\subsection{Learning Objectives and Optimization}

\mysection{Training Label Collection}
To collect high-quality labels for training, we use customer engagement behaviors within a certain attribution window starting from item impressions on search results.
This delay is crucial for correctly attributing conversions that are not immediate. For example, a customer might add an item to their cart but only complete the purchase a day later; the attribution approach ensures that this delayed purchase is successfully linked back to the original search impression.

\mysection{Multi-task Learning}
The target conversion behaviors (e.g. purchases in this work) are usually extremely sparse compared to total exposures.
To enrich learning signals and smooth the learning process, it is common and useful to take advantage of other engagement behaviors, such as clicks and add-to-carts~\cite{zhuang2025practice}.
Hence, it is natural to apply multi-task learning to simultaneously learn to predict different behaviors.

For each learning task $t$, we use the Softmax cross-entropy loss~\cite{bruch2019analysis} for optimization as:
$$\mathcal{L}_t  = -\sum_i y_{i,t} \log \frac{\exp\left(s_{i,t}\right)}{\sum_j \exp\left(s_{j,t}\right)},$$
where $y_{i,t}$ denotes if the item $d_i$ results in the engagement behavior for the task $t$; $s_{i,t}$ is the predicted logit on the task $t$ for the item $d_i$.
The overall learning objective can then be simply calculated as the summation of the losses of all learning tasks.
For optimization, in this work, we use Adam optimizer~\cite{kingma2014adam} with a portion of the first epoch for learning rate warm-up and use a cosine decay until the end of training~\cite{devlin2019bert}.

% \jj{Do we want to reveal the exact behaviors (multi-head) in multi-task learnings or just mention we use a few engagement behaviors and keep it private?}

\if 0
\subsection{training data}
When a customer interacts with the search results page (SRP), two distinct data streams are generated. First, impression logs capture the search results presented to the customer, including the specific features and their versions at the time of the query. Second, customer actions, such as clicks or purchases, are recorded in separate interaction logs.

To construct the final training dataset, the interaction logs are joined with the impression data along with additional features after a predefined attribution window. This delay is crucial for correctly attributing conversions that are not immediate. For example, a customer might add an item to their cart but only complete the purchase a day later; the attribution window ensures this delayed purchase is successfully linked back to the original search impression.
\fi

%% file: contents/section3_efficientmodeling.tex
For model scaling, we hypothesize that we can achieve considerable quality gains with more computation, more model parameters, and more training data.
 However, scaling up models also typically leads to higher cost and more demanding hardware requirements.
Therefore, how to predictably but efficiently scale up models becomes a difficult challenge.
In this work, we explore the following three strategies to scale up CVR models:

\begin{itemize}[leftmargin=*]
   \item \textbf{Backbone Scaling} increases the computational intensity in the backbone of the feature interaction module. This typically reduces training throughput as the floating-point operations (FLOPs) per sample increase.
   \item \textbf{Embedding Scaling} increases the number of model parameters by expanding the dimension or quantity of embedding tables. While often computationally efficient, this significantly grows the memory footprint, which can trigger memory bottlenecks. 
  \item \textbf{Data Scaling} involves training on a larger corpus by extending the data duration or increasing the sampling rate. This results in longer training cycles that may become infeasible under strict training budgets.
\end{itemize}

\if 0
\subsubsection{For non-sequential models} ~\\
For non-sequential architectures like MLPs and MaskNet, the 2D tensor input ($B \times D$) 
is formed by concatenating all features into a $D$-dimensional vector, 
after which Batch Normalization is applied to standardize the inputs. This 
approach is straightforward and standard for models not designed to process sequential data.

\subsubsection{For sequential models}  ~\\
Sequential models like the Transformer or RankMixer require a 3D tensor input 
($B \times S \times D_{\text{emb}}$), posing a significant challenge 
in structuring heterogeneous features into a meaningful sequence of tokens.
\begin{itemize}
    \item \textbf{Conventional Method:} Relies on manual, heuristic-based 
    grouping of features, where each group is projected into a token. 
    This process is not scalable and is inflexible to changes in the 
    feature set.
    
    \item \textbf{Proposed Method:} We utilize a Multi-Layer Perceptron (MLP) 
    to automatically learn the feature groupings and token representations 
    from data. This data-driven approach automates the creation of the 3D 
    tensor, eliminating the need for manual engineering and resulting in a 
    more scalable and adaptable input pipeline.
\end{itemize}
\fi 

\subsection{Backbone Scaling}
\label{section:backbone_scaling}

%The amount of leveraged computation is known to be one of the most important factors for machine learning models to be more effective~\cite{sutton2019bitter}.
%In modern CVR models, most computations are determined by the selection and configuration of the backbone models in the feature interaction module.
The effectiveness of machine learning models is significantly driven by their ability to leverage additional computational capacity~\cite{sutton2019bitter}. In modern CVR prediction, this capacity is largely dictated by the architecture and configuration of the backbone models within the feature interaction module.
In this paper, we explore two directions: (1) scaling individual backbone models across different scaling factors and (2) scaling the computation architecture for feature interaction by ensembling heterogeneous backbones.

\mysection{Individual Backbone Scaling}
For the selection of backbone models, we consider several conventional state-of-the-art models including 
DCNv2~\cite{wang2021dcnv2}, MaskNet~\cite{wang2021masknet}, Transformers~\cite{vaswani2017attention} and RankMixer~\cite{zhu2025rankmixer}. %These model architecture options are consistent with prior research~\cite{zhuang2025practice}. 
We begin by discussing the model architecture family and its associated scaling factors. 

\mysection{The Deep and Cross Family}
The family of models aims to combine standard fully connected deep neural network layers (FC-DNN) with Hadamard-product-based cross-feature interaction layers. We explore {\bf DCNv2} and {\bf MaskNet}.
Although both models are conceptually similar, they differ in their cross-interaction logic. DCNv2 computes
$$\mathbf{x}_{l+1} = \mathbf{x}_0 \odot (\mathbf{U}_l\mathbf{V}_l \mathbf{x}_l) + \mathbf{x}_l,$$
whereas MaskNet uses
$$\mathbf{x}_{l+1} = \mathbf{U}_l(\text{ReLU}(\mathbf{V}_l\mathbf{x}_0)) \odot \mathbf{x}_l.$$ Bias terms are omitted for clarity. To enhance the expressive power of these cross layers in an unbounded way, we propose to scale the input $\mathbf{x}_0$ via one-layer relu projection, addressing the bottleneck in DCNv2 where interaction rank is limited by the input vector length.  

Regarding FC-DNN connectivity, MaskNet positions the DNN immediately following the cross-network output. It utilizes $n$ MaskNet blocks, $M_i(\mathbf{x}, \mathbf{x}_0)$, whose outputs can be concatenated in parallel as:
$$\text{Concat}(M_0(\mathbf{x_0}, \mathbf{x}_0), M_1(\mathbf{x_0}, \mathbf{x}_0), \cdots, M_n(\mathbf{x_0}, \mathbf{x}_0));$$
or stacked sequentially as:
$$M_n( \dots (M_2(M_1(\mathbf{x_0}, \mathbf{x_0}), \mathbf{x_0}), \dots, \mathbf{x_0})$$
Following these operations, the resulting hidden representation is projected directly to the output logit for prediction. This architectural flexibility mirrors DCNv2, which permits the FC-DNN to be either stacked after the cross-layer output or integrated in parallel via concatenation.

To evaluate the impact of these architectural choices, we identify four key scaling factors:
\begin{itemize}%[leftmargin=*]
\item Cross Width: the projected cross-input dimension
\item Deep Width: the FC-DNN hidden dimension
\item Parallel Mask Blocks: number of MaskNet parallel blocks
\item Sequential Mask Blocks: number of MaskNet stacked blocks
\end{itemize}

\mysection{The Sequence Model Family} Modern CVR models increasingly leverage sequence embeddings to unlock the power of architectures like Transformers and RankMixer. While Transformers excel at capturing high-order, contextual interactions through self-attention, RankMixer enhances scalability by replacing quadratic attention with efficient multi-head token mixing.

To integrate non-sequential signals, such as 3-month aggregate conversion rates, we adopt the established practice of concatenating and uniformly partitioning features into a sequence of ``global tokens.''\cite{xu2025climber, zhu2025rankmixer} To enhance representational power, we apply a non-linear projection to raw features prior to splitting, mirroring techniques used in Cross-Net architectures.

To assess the impact of different architectural choices, we identify five key scaling factors:
\begin{itemize}%[leftmargin=*]
\item $d_{model}$:  the size of the input embedding
\item Sequence Length: the number of sequential ``global tokens''
\item $\text{ffn}_{\text{dim}}$: the dimension size in the ffn layers
\item The number of model layers
\item The number of attention heads
\end{itemize}

\mysection{Backbone Model Ensemble}
To properly derive an ensemble of backbone models into a CVR model, we use deep hierarchical ensemble network (DHEN)~\cite{zhang2022dhen,zhuang2025practice}. We aim to understand how the improvements add up for backbone combinations with different ensemble approaches.

%\todo{Revisit the method about DHEN ensemble when having conclusions.}

%Note that from the model perspective, compute scaling discussed in this section only refers to the increase in the computational amount and could be orthogonally independent of the deployment of more computational resources discussed in Section~\ref{section:infra}.

\subsection{Embedding Scaling}

Modern conversion prediction and recommendation models rely on embedding techniques to handle categorical features with high cardinality item IDs and text tokens.
These trainable embedding tables usually account for a meaningful portion of the model parameters~\cite{DLRM19}.
%Accordingly, a natural way to scale up models is to further enlarge these embedding tables, thereby introducing more degrees of freedom in these continuous representations.
Accordingly, a natural way to scale up models is expanding these tables in order to enhance the representational capacity of the model.

In this work, we explore two factors for embedding scaling: (1) the size of embedding dimensions and (2) the vocabulary size.

While wider dimensions improve the richness of the continuous representations, extending the vocabulary allows the model to distinguish between a broader range of features. To manage the massive cardinality of features like n-grams, we rely on a hashing trick~\cite{kang2021learning}, which allows precise control over vocabulary size by adjusting the hashing modulus divisor.

%%[VJ] do we need these remarks? move closer to Table 1?

%%most ranking models have multiple embeddings for item id features, token id features and categorical features. We explore increasing the embedding table size (vocab size or embedding size) or increasing the number of embedding tables used without adding major computation. 

%%For parameter scaling, we mainly look at trainable embeddings inside the model. For our model, there are token embeddings for query and item (including title, brand, etc.) tokens. There are embeddings from categorical features. Lastly, there are product-id embeddings. Usually, the product set is of high cardinality, and we hash the product id to 0.1\% of the product cardinality vocab set. 

%%We focus on the product id embeddings are they are the most powerful embedding in our past experience. A reason is that product ids are used in many places, at a feature itself, as well as used in the past engaged product id sequence for customer-historical sequence features. 

%%When we scale, we can scale the vocab size (covering more of that 0.1\%) or embedding size (making the embedding larger).

\subsection{Data Scaling}
\label{section:datascaling}

Learning from historical engagement behaviors is a standard approach in CVR modeling. From a learning signal perspective, scaling training data by increasing the sampling rate or time duration enhance model quality and generalizability~\cite{zhang2024wukong}.
Larger datasets provide more comprehensive coverage of customer behaviors, particularly for torso and tail search queries where engagement is inherently sparse. Extended training windows enable the model to capture diverse temporal patterns and evolving user interests.

In production, feature sets evolve dynamically; newer data partitions tend to exhibit superior feature coverage as additional signals are integrated. While generating all features across the full training lookback period is ideal, it is often infeasible. Thus, we employ a technique of populating missing features with a unique `unseen' placeholder, evaluating whether the substantial cost of extensive backfilling yields commensurate performance gains. Recognizing that data 
scaling is not a “free lunch,” we prioritize targeted feature 
exploration, backfilling only a scoped subset of high-impact signals.

%% file: contents/section4_exp.tex
%In this section, we conduct extensive experiments and in-depth analysis to verify our scaling strategies, proposed hypotheses, and online serving optimization.
%Online A/B tests also demonstrate consistent improvements in production.

%\subsection{Experimental Settings}

\input{tables/compute-scaling-performance}
\mysection{Experimental Datasets}
For experiments, we used 70 days of anonymized purchase data for model training, while more or less days of training data are also considered in the study of data scaling.
Each training sample is a search result list of items with at least one purchase.
We consider purchase engagements as positive labels for the learning-to-rank softmax loss~\cite{bruch2019analysis} to train the model.

\mysection{Backbone Models and Features}
As mentioned in Section~\ref{section:efficientmodeling}, we use DCNv2~\cite{wang2021dcnv2}, MaskNet~\cite{wang2021masknet}, Transformers~\cite{vaswani2017attention}, and RankMixer~\cite{zhu2025rankmixer} as backbone models.
%We use a parallel DCNv2 model~\cite{wang2021dcnv2} with hidden sizes of $\left[768, 768, 768\right]$ for the deep network part and a low-rank cross size of 96 with one cross-layer, followed by a feature projection with a 675-dimensional hidden layer.
%The model also includes four DCNv2 experts controlled by query statistics features, so different queries can activate corresponding experts.
Our models utilize various features that include query-item engagement features, query engagement features, item engagement features, query-item understanding features and customers' past engagement sequences. The models also utilizes raw query, item title, and document ID features, with their embeddings learned jointly during training.
%Table~\ref{tab:feature_counts} shows the percentage distribution of different feature groups. Note that as part of customer engagement features, we also include some sequence features.

%%[VJ] Why do we need Table3? Feels random. Also I made a note "that as part of customer engagement features, we also include some sequence features" because it made it seem like we didn't include any sequence features.
 
%\input{tables/features}

\mysection{Evaluation Metrics}
For evaluation, we compute the mean average precision (mAP) of all searches to customer purchases for seven days in the future because we have found that this metric correlates well with online customer engagement. 
Formally, mAP in our experiments can be derived as:
$\text{mAP} = \frac{1}{Q} \sum_{q=1}^{Q} \text{AP}_q,$
where 
$$\text{AP}_q = \frac{\sum_{k=1}^{n} (\text{precision}@k \times \text{item purchased}@k)}{\text{Number of purchased items for query } q}.$$
Note that because most queries result in a single purchase, the value of this metric closely approximates the Mean Reciprocal Rank (MRR).
In our experiments, we will report the relative offline metric difference (\% of mAP) of model changes.

%, where a 0.1\% mAP gain threshold is considered significant. 
%%[VJ] added "a 0.1\% mAP gain is considered significant. "

 \if 0
 
 After getting the features, we do a batch normalization and project the feature into a 675 dimensional hidden layer. We have four DCNv2 experts with weight controlled by some query statistic features with a idea that different queries would trigger different experts. The model has $O(600)$ features, including query-item engagement features, query engagement feature, item engagement features, raw query, raw item features, query/item understanding features and personalization features.  See Table~\ref{tab:feature_counts} for a breakdown of distribution of such features. It is important to understand our study is based on the current feature set. Some previous work has found that engagement features may be replaced by long sequence models with raw engagement actions. Scaling laws may be different in those scenarios. We will leave it for future study.
\fi

\if 0

We conduct experiments to study the following research questions to help us navigate scaling better.

\begin{itemize}
  \item RQ1 - Do some models scale better than others?
  \item RQ2 - does more data help model scaling?
  \item RQ3 - how different gains transfer into the final model?
  \item RQ4 - do different feature groups have different feature importance after scaling?
\end{itemize}
\fi

\input{tables/backbone-scaling-exp}

\subsection{Results on Backbone Scaling}
To evaluate backbone scaling, we comprehensively scale each model along every scaling factor highlighted in Section~\ref{section:backbone_scaling}.
Table~\ref{tab:backbone_scaling_exp} presents the relative changes observed when utilizing different backbones in the feature interaction module.

\begin{figure*}[!t]
    \centering
    \includegraphics[width=1\textwidth]{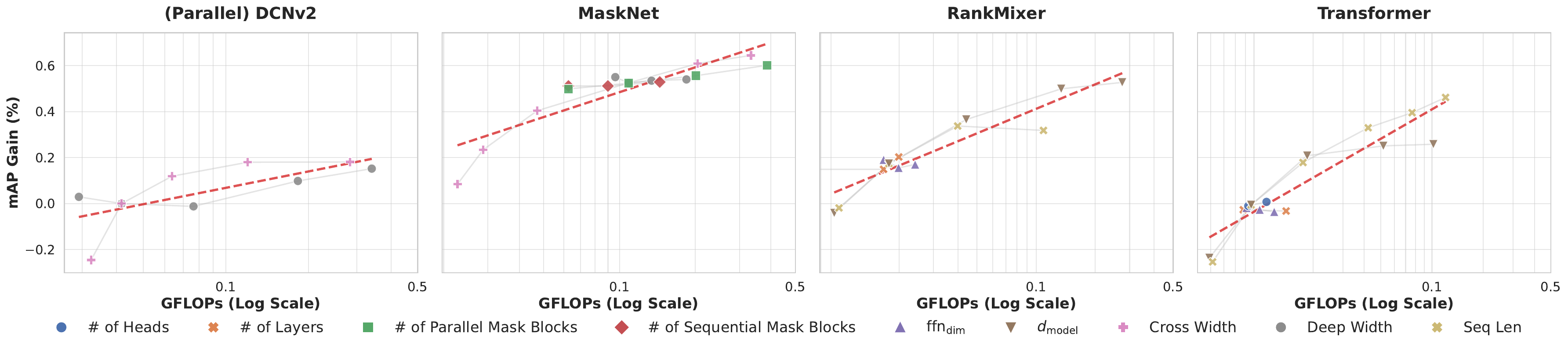}
    \caption{FLOPs vs mAP gain against DCNv2 for different scaling factors.}
    \label{fig:sizevsmap}
    %\vspace{-12pt}
\end{figure*}

\mysection{Model Backbone}
Figure~\ref{fig:sizevsmap} compares mAP across several backbone models with more data points.
Initially, we observed that sequence models outperform DCNv2 within a similar FLOPs budget. This aligns with prior findings for models of this scale (about $10^8$ FLOPs) \cite{xu2025climber}, which noted that Transformers generally outperform DLRM-family models.
However, our well-tuned MaskNet exceeds the performance of sequence models.
We conjecture that MaskNet outperforms sequence models with attention and rank-mixing by more effectively capturing feature interactions through its cross-scaling mechanism.Regarding hardware efficiency, Table~\ref{tab:backbone_scaling_exp} shows that MaskNet V4 variants incur a 29\% throughput drop at $2.5 \times 10^8$ FLOPs. However, sequence models yield similar or lower throughput at only half that computational load, suggesting that MaskNet is significantly more efficient per FLOP.

Beyond throughput, we observed critical distinctions in memory consumption across architectures.
Despite having comparable FLOP counts, sequence models exhibit disproportionate memory overhead, which impedes scaling beyond $6\times$ the baseline factor values (see Table~\ref{tab:backbone_scaling_exp}).
This is primarily attributable to the memory required to store hidden states per timestamp and attention matrices for backpropagation, which leads to out-of-memory (OOM) errors significantly earlier than their deep-and-cross counterparts. The memory bottleneck becomes more pronounced in deeper sequence models. Overall, we aim to highlight that selecting the right backbone considering performance and efficiency is critical for production viability.

\mysection{Scaling Factors} Selecting the right scaling factors is also essential: for example, we find scaling 
{\it cross-width} scaling is more effective than {\it deep-width} scaling. 
Similarly, for Transformers, scaling the sequence length emerges as the most 
effective factor. We hypothesize that the superiority of these specific 
axes stems from the nature of search and recommendation tasks, which 
prioritize explicit high-order feature interactions. This capability is 
enhanced by expanding cross-layers in MaskNets and sequence 
capacity in Transformers, as both facilitate more exhaustive feature crossing.

To investigate this behavior, we conducted a permutation feature importance analysis, measuring mAP fluctuations after shuffling individual features and feature pairs across the top-performing model variants. While first-order feature importance remains consistent across all models, second-order interactions involving customer embeddings exhibit significant variance. As shown in Table~\ref{table:feature_importance}, higher mAP score models correlate with a second-order feature importance distribution that closely aligns with that of MaskNet. This may suggest that the ability to accurately learn feature crosses is a primary driver of superior predictive performance.

\begin{table}[!t]
\centering
\begin{tabular}{c|cc}
\hline
 & \multicolumn{2}{c}{\bf Feature Category Importance} \\ \cline{2-3}
\textbf{Model Type} & \textbf{Engagement} & \textbf{Item/Query Understanding} \\ \hline
MaskNet     & 47.90\% & 52.10\% \\
RankMixer   & 46.14\% & 53.86\% \\ 
Transformer & 44.03\% & 55.97\% \\
DCNv2       & 42.36\% & 57.64\% \\ \hline
\end{tabular}
\caption{Normalized second-order feature importance (interactions between customer embeddings and other feature categories) for the top-performing variant of each backbone. Note the mAP ranking of the top variant from Table~\ref{tab:backbone_scaling_exp} also follow this order: MaskNet, RankMixer, Transformer, and finally DCNv2.}
\label{table:feature_importance}
\end{table}

\mysection{Scaling Properties} Consistent with scaling laws observed in LLMs and prior research, we observe log-linear improvements across backbones and most scaling factors. However, as certain axes demonstrate greater scaling efficiency compared to others, identifying which factors to scale is as vital as initial backbone selection.

% Following Hypothesis~\ref{hypothesis:indscale}, we combine individual scaling strategies and further scale up the model by ensembling different model backbones.
% %Table~\ref{tab:stack_model_scaling_performance} demonstrates the overall improvements with different ensembles of models.
% Note that because of hardware and resource limitations, we do not combine 8x models or combine more than three backbones in an ensemble.
% One thing to observe is that Transformer-4x and RankMixer-4x ensemble results in a similar improvement to the individual DCNv2-4x with a smaller increase in the model size. In addition, combining three different backbones, Transformer-4x, RankMixer-4x, and MaskNet-4x, can achieve the best performance, albeit a marginal improvement over the simpler two-backbone ensemble.
% However, not all ensembles that result in an increase in model size improve performance.

\subsection{Results on Embedding Scaling}

For embedding scaling, we measure the mAP increase as we scale the embedding vocab size and dimension of different embeddings of item IDs, queries, and product titles.

For item embedding scaling, we explored increasing both the embedding dimension and the vocabulary size. To scale the vocabulary, we simply increased the hash modulo, as our system utilizes item ID hashing. As shown in Table \ref{tab:iid_embedding_vocab_size}, scaling the embedding dimension proved significantly more effective than increasing the vocabulary size. Specifically, doubling the embedding size yielded greater performance gains than a 10x increase in vocabulary size. We hypothesize that this is due to the heavy concentration of user interactions on a small subset of ``top items.'' Furthermore, we observed that the gains from these two scaling dimensions are largely additive, suggesting they capture complementary information.

\input{tables/iid-embedding-table}

For queries and product titles, we focused on scaling the dimension size of token embeddings as the basis for text embeddings with a pooling layer.
Our findings indicate that doubling the query embedding size yields a gain of 0.03\% mAP, after which the performance plateaus.
For product titles, embedding scaling provides no gain at all. 
We hypothesize that scaling text embeddings does not work as categorical features like item IDs because individual text tokens need to be shared and universal for general texts.
Compared to categorical features with very specific semantics, each text token could be more general in describing various knowledge with other tokens, so it is limited to scale the embedding dimension of tokens when the key is the mixture of different text tokens.

% TODO(Fan): do {1M, 10M} x {16d, 32d, 64d, 100d}. to understand if larger vocab (intuitively) can raise the ceiling of embedding dimension. 

\if 0
{\it 
In previous investigations, we have established more data and more compute helps with mode quality as long as we scale them efficiently. 

In reality, we do not only scale one dimension at a time, but we scale multiple dimensions in parallel. A natural question is about how we combine multiple good changes together. And once we combine, how do the result look like. Here, for modeling changes, we combine them in parallel in our DHEN framework since our DHEN is one layer only, it means we just add a parallel component next to an existing network.

From Table~\ref{tab:model_metrics} we can see that the gains to mostly stack on top of each other without a major haircut. To be more specific, Masknet + wider DNN has a gain of 0.43\% which is not far from MaskNet improvement of 0.34\% plus DNN improvement of 0.13\%.
}
\fi 

\if 0
\begin{figure}
    \centering
    \resizebox{\linewidth}{!}{
        \input{figures/samplingaccumulation}
    }
    \caption{Caption TK}
    \label{fig:placeholder}
\end{figure}
\fi

\subsection{Results on Data Scaling}
\begin{figure}[!t]
    \centering
    \includegraphics[width=.8\linewidth]{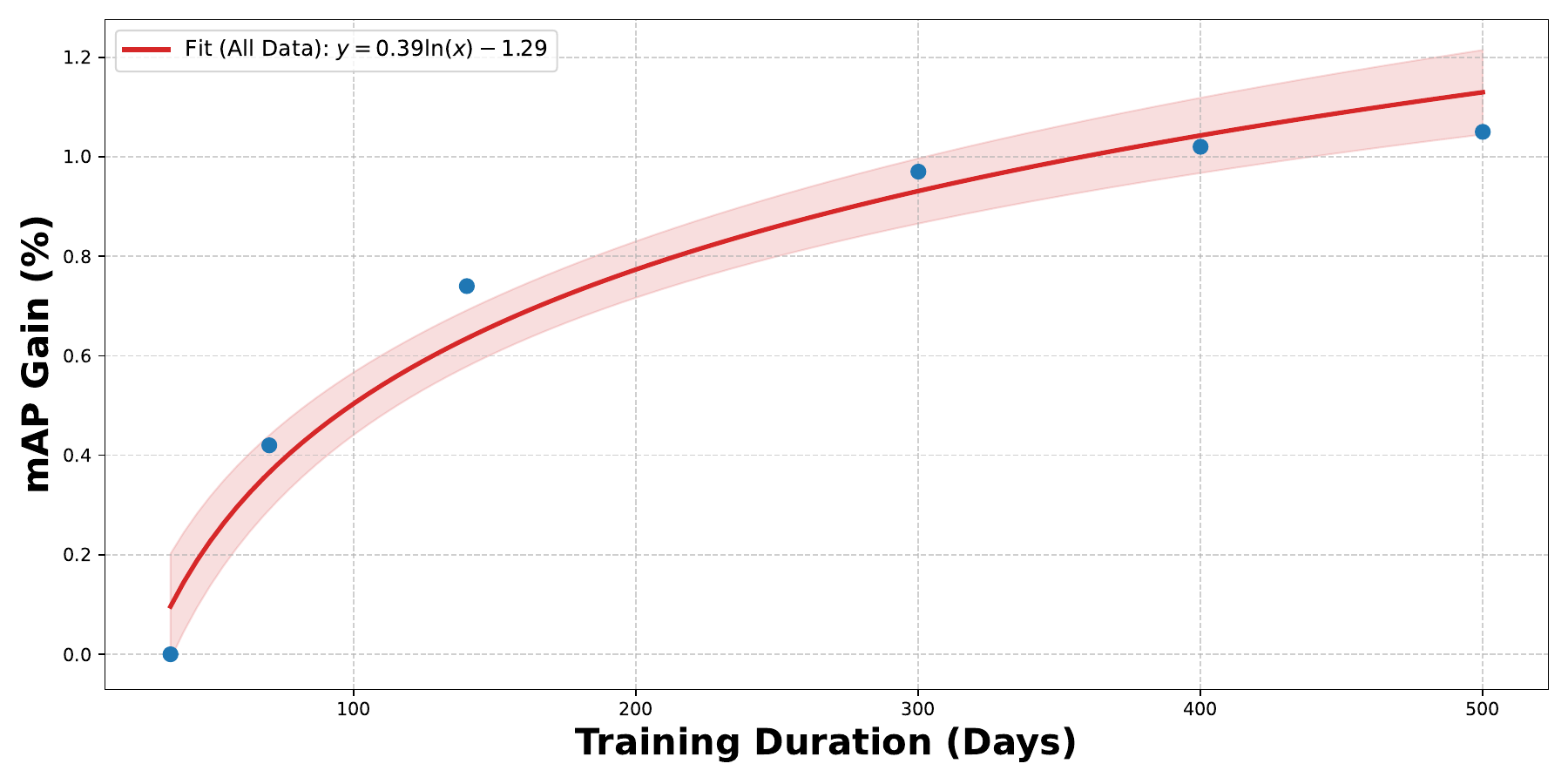}
    \caption{Impact of  data scaling for a MaskNet model on mAP.}
    \label{fig:data_scaling}
    %\vspace{-21pt}
\end{figure}

Figure~\ref{fig:data_scaling} illustrates that the mAP gain across varying training data volumes follows a predictable log-linear scaling law. Because training cost and time increase linearly with 
data volume, the diminishing returns visible in the flattening curve make 
simple volume scaling highly cost-ineffective for production environments. We mitigate this training efficiency bottleneck through 
our streamlined warmstart strategy, detailed in Section~\ref{section:online_training_serving_optimization}. In addition to diminishing marginal returns, a longer training window also introduces a feature gap, as historical 
logs usually lack newly developed signals. To assess 
feature availability, we compared a full feature set against a configuration 
where personalization embeddings were restricted to the most recent 
300 days.
Note that personalization embeddings only consider historical item engagements as an input sequence and do not involve any personally identifiable information.
As shown in Table~\ref{tab:missing_features_data_scaling}, this 
coverage gap did not meaningfully degrade performance. This key 
observation facilitates a more targeted feature exploration strategy, 
allowing us to prioritize high-signal, recent data for new features without 
the need for expensive historical backfilling or simulation.

\input{tables/missing-features-data-scaling}

\begin{figure}[!h]
    \centering
    \includegraphics[width=0.8\linewidth]{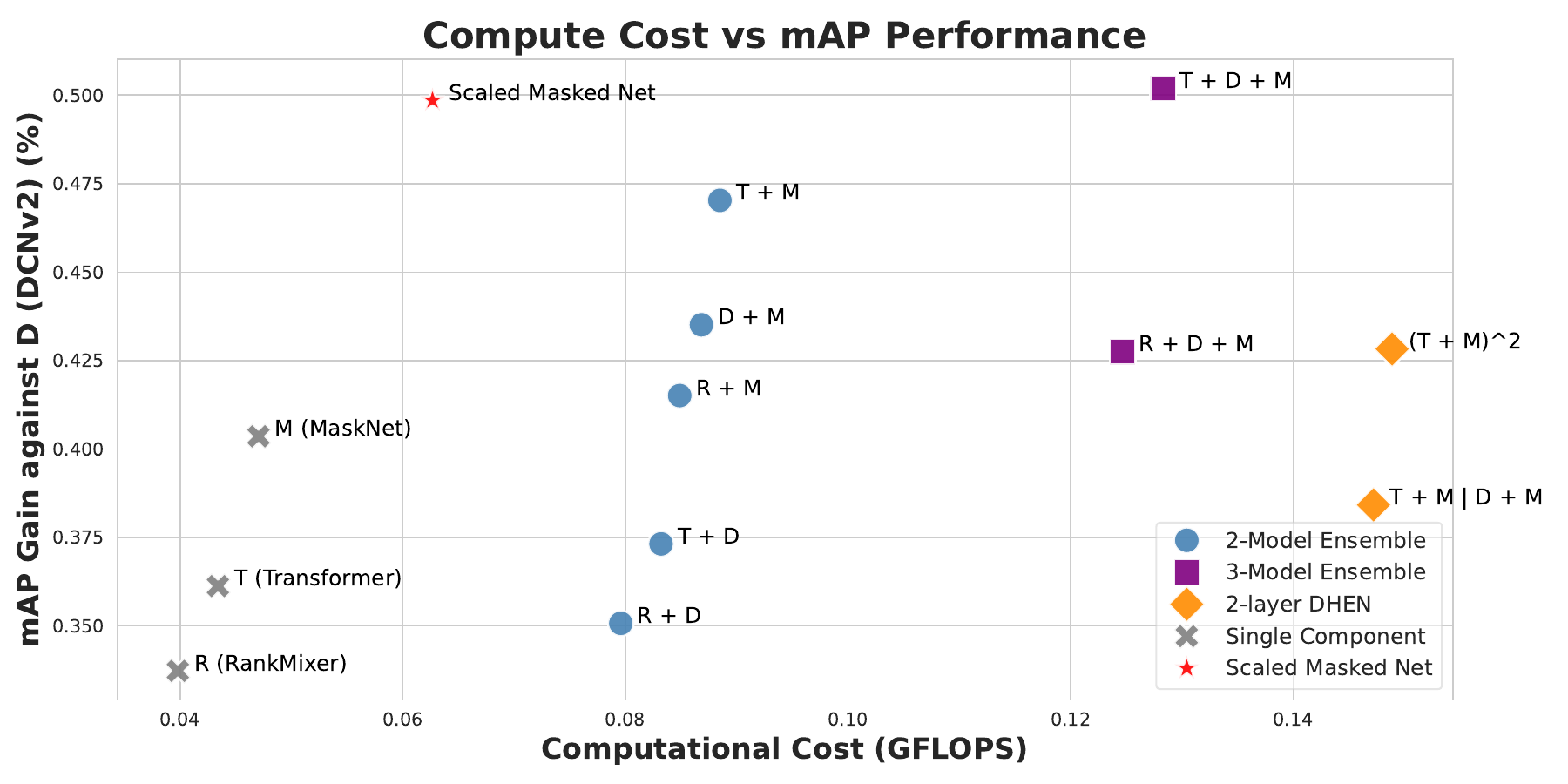}
    \caption{FLOPs vs mAP gain for models and scaling factors.}
    \label{fig:ensemble}
    \vspace{-6pt}
\end{figure}

\subsection{The Compound Effect of Scaling}
We look at how combining different model backbones and scaling up data, compute, and embeddings affects performance. Crucially, we test if these benefits add up when used together. Proving that these methods can be used independently allows us to optimize each one separately, making our overall development much faster.

\mysection{Backbone Ensemble}
We explore DHEN~\cite{zhuang2025practice,zhang2022dhen} as a scaling strategy with multiple high-performance backbone models. % when multiple high-performing backbone models are available.
As shown in Figure~\ref{fig:ensemble}, while single-layer model ensembles consistently outperform individual components in terms of mAP, they are ultimately surpassed by our optimized single-model architectures when evaluating the mAP-compute tradeoff.
The consistent mAP improvements of DHENs make it a reliable choice for immediate production gains \cite{zhuang2025practice}. However, proper scaling a single backbone is the more sustainable path toward long-term model Pareto-optimality.

% \mysection{Backbone + Data Scaling}
% Figure~\ref{fig:datascaling} shows the mAP gain against DCNv2 for several model backbones with different sizes of training data.
% Notably, our analysis indicates that the relative performance remains consistent across the number of training days.
% This enables a more efficient research trajectory, as architectural refinements can be validated on smaller datasets while maintaining generalizability to full-scale training.
%It also implies that the benefits of scaling data and the backbone are additive

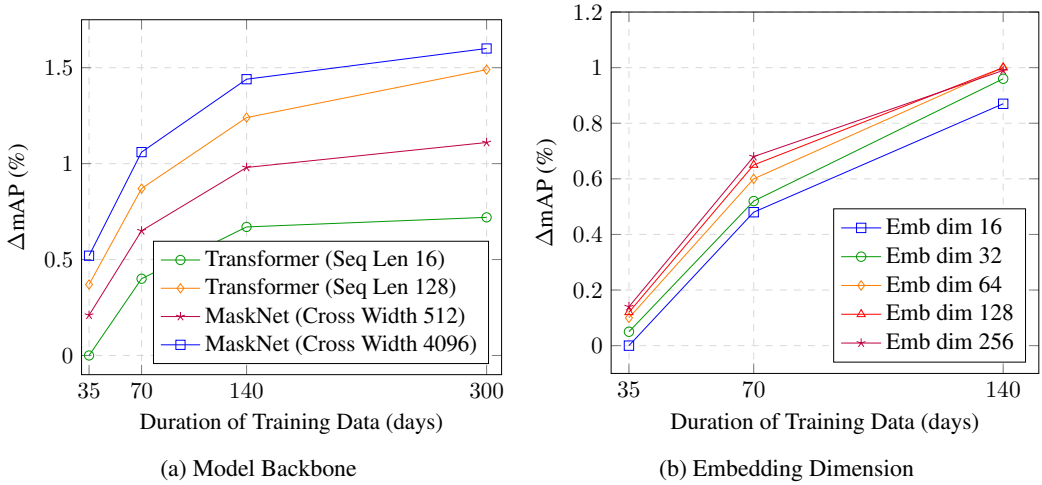
\begin{figure}[!h]
    \centering
    \begin{subfigure}[b]{0.5\linewidth}
        \centering
        \resizebox{\linewidth}{!}{
        \input{figures/data-scaling-line-chart}
        }
        \caption{Model Backbone}
        \label{fig:datascaling}
    \end{subfigure}%
    \begin{subfigure}[b]{0.5\linewidth}
        \centering
        \resizebox{\linewidth}{!}{
        \input{figures/emb_data_scaling_line_chart_v2}
        }
        \caption{Embedding Dimension}
        \label{fig:embdatascaling}
    \end{subfigure}%
    \caption{Performance gains over different durations of training data with (a) different model backbones and with (b) different item embedding dimension sizes for MaskNet.}
    \label{fig:datascalingmerged}
    %\vspace{-6pt}
\end{figure}

\mysection{Backbone/Embedding + Data Scaling}
Figure~\ref{fig:datascalingmerged} illustrates the mAP gain with different sizes of training data on different model backbones and sizes of item embedding dimension. 
%Figure~\ref{fig:datascaling} shows the mAP gain against DCNv2 for model backbones with different sizes of training data.
%Figure~\ref{fig:embdatascaling} illustrates the mAP gain as we scale the embedding with different sizes of training data for MaskNet.
Our analysis indicates that relative performance remains consistent across different sizes of training data.
It enables a more efficient research trajectory, as architectural refinements can be validated on smaller datasets while maintaining generalizability to full-scale training.

\if 0
\mysection{Embedding + Data Scaling}
Notably, our analysis indicates that the relative performance remains consistent across the number of training days.
This enables a more efficient research trajectory, as architectural refinements can be validated on smaller datasets while maintaining generalizability to full-scale training.
\fi 

% \begin{figure}[!t]
%     \centering
%     \resizebox{\linewidth}{!}{
%         \input{figures/emb_data_scaling_line_chart}
%     }
%     \caption{Performance gains of different item id embedding dimensions based on 1M vocab size over different durations of training data. %TODO: Data to be updated
%     }
%     \label{fig:embdatascaling}
%     %\vspace{-20pt}
% \end{figure}
\if 0
\begin{figure}[!t]
    \centering
    \resizebox{\linewidth}{!}{
        \input{figures/emb_data_scaling_line_chart_v2}
    }
    \caption{Performance gains of different item id embedding dimensions based on 1M vocab size over different durations of training data. %TODO: Data to be updated
    }
    \label{fig:embdatascaling}
    %\vspace{-20pt}
\end{figure}
\fi

\mysection{All Scaling Combined}
Table~\ref{tab:model_gains} summarizes the mAP improvements relative to the baseline. We observe that scaling the training duration, backbone, and embedding size together achieves a 0.74\% improvement, demonstrating that these scaling effects are complementary and largely additive, allowing us to experiment with each independently before combining them.

\if 0
\subsubsection{Data Scaling + Model Scaling}
In this section, we analyze how different amounts of data would affect the performance, as well as data scaling strategies during exploration. We also study the potential combined effect of data scaling plus model scaling.

\subsubsection{Effective Data Scaling}

While larger datasets typically yield better mAP, the training cost increases linearly with dataset size. To address this trade-off, we adopt an intuitive sampling strategy based on two principles: recent data is more important, and data from consecutive days is often redundant. We propose an algorithm that, when selecting $N$ days from a pool of $M$ (where $N < M$), combines recency with diversity. It selects the $N/2$ most recent days and then uniformly samples the remaining $N/2$ days from the older $M - N/2$ days.

In an experiment with $M=240$ total days and a target of $N=120$ training days, our approach achieved a 0.24\% mAP gain over a baseline using only the 120 most recent days, with no additional training cost. For comparison, training on all 240 days yielded a slightly higher 0.27\% gain but required double the training time. 

If we just do uniform sampling of $N$ days out of $M$ days, in this case, we only get 0.14\% gain. The hybrid approach is needed.

\subsection{Scaling vs. Efficiency}

TK: update this section once we confirm a few representative models

\begin{itemize}

    \item latency vs flops comparison between models

    \item embedding table vs dense models

\end{itemize}

\fi

\begin{table}[!h]
\centering
\begin{tabular}{l|c}
\hline
\multirow{2}{*}{\textbf{Scaling Dimension}} & \multicolumn{1}{c}{\bf Model Scaling Impact} \\ \cline{2-2}
 & \textbf{mAP Gains (\%)} \\ \hline
Data     & +0.47\% \\
Backbone & +0.10\% \\
Embedding       & +0.13\% \\ \hline
\textbf{Combined} & \textbf{+0.74\%} \\ \hline
\end{tabular}
\caption{Incremental and cumulative performance gains over a pre-scaling MaskNet model. The combined model integrates all scaling optimizations.}
\label{tab:model_gains}
%\vspace{-24pt}
\end{table}

\subsection{Training and Serving Optimization}
\label{section:online_training_serving_optimization}

Our optimization strategy addresses two primary constraints: serving latency and training iteration speed. We prioritize several key initiatives—training warmstarts, decoupled CPU-GPU execution, and dynamic batching—while further reducing compute overhead through continuous feature pruning and moving feature preprocessing outside model execution.

\mysection{Streamlined Warmstart for Faster Training Iterations} Rapid online A/B testing necessitates frequent offline model iterations, making training efficiency critical. To accelerate convergence and reduce resource consumption, we implemented a warmstart strategy. We aim to streamline training iterations without sacrificing model quality. We first train a high-capacity base model (e.g. one year of data with an 8x backbone) and subsequently fine-tune the entire architecture on the most recent 140 days. A primary challenge in warmstarting is that feature changes often alter the input dimensionality, rendering previous checkpoints incompatible. Our design mitigates this by projecting the input state into a fixed-dimensional hidden layer as shown in Section~\ref{section:backbone_scaling}. This allows us to initiate training from existing models without concern for internal parameter compatibility or manual coordination. 
When features are added or removed, we re-initialize only the initial projection weights while warming all subsequent parameters simultaneously. This unified update philosophy (i.e. warming all available layers at once and updating every parameter) ensures that the fine-tuning process is as seamless as training from scratch. Our evaluation demonstrates that this strategy has a negligible impact on mAP, meaning a fine-tuned model matches the quality of one trained from scratch. This facilitates rapid feature iteration and maintains high model quality while substantially reducing training overhead. To prevent cumulative drift and maintain reproducibility, we re-train the base model on a periodic cadence or after the launch of significant model improvements.

\begin{figure}[!h]
    \centering
    \resizebox{.8\linewidth}{!}{
    \input{figures/cpu-vs-gpu-tikz}
    }
    \caption{P99 latency of CPU serving, as well as unoptimized and optimized GPU serving with hybrid CPU-GPU execution.} 
    
    %Model inference optimization with workload-aware device placement.
    %CPU: Latency on CPU rises quickly as we increase batch size of a request.
    %GPU (baseline): Latency on GPU is slower than CPU at smaller batch size due to CPU-GPU communication.
    %GPU (optimized): Reduced p99 latency of batch size=600 from 82ms to 32ms by placing input tensors, feature transformation, and embedding lookup on CPU.
    %}
    \label{fig:hardware-comparison-for-model-inference}
    %\vspace{-9pt}
\end{figure}
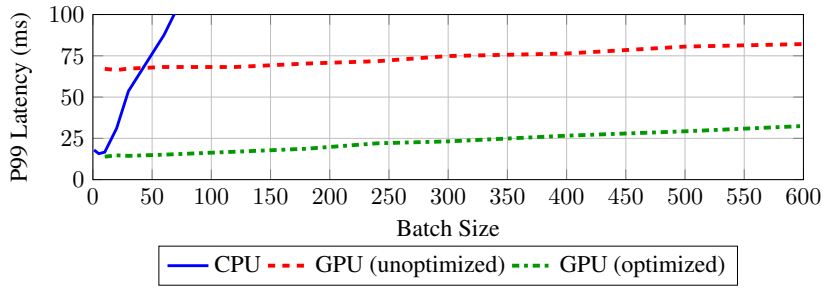

\mysection{Inference Performance Optimization} We optimized our serving architecture to meet latency SLOs under dynamic load. \begin{itemize}[leftmargin=*] \item \textbf{Hybrid CPU-GPU Execution:} Initial profiling revealed GPU communication bottlenecks, making CPU-only inference faster. By splitting the graph—pinning preprocessing to the CPU and the backbone to the GPU, we reduced the P99 latency by 2.6x from 82ms to 32ms as shown in Figure~\ref{fig:hardware-comparison-for-model-inference}. \item \textbf{Dynamic Batching Strategy:} To maximize GPU saturation, we employed a dual-batching approach. \textit{Client-Side Batching} groups candidate items upstream, allowing GPUs to process 20x larger batches with minimal latency overhead as shown in Figure~\ref{fig:hardware-comparison-for-model-inference}. For high-concurrency traffic, \textit{Server-Side Batching} dynamically aggregates independent requests. By tuning the batch timeout to an optimal 10ms, we increased peak throughput by 4.4x as shown in Table~\ref{tab:server_side_batching} while maintaining acceptable latency limits. \end{itemize}

% \input{tables/client-side-batching} 
\input{tables/server-side-batching}

% \subsection{Online Performance}

% In addition to offline experiments, we also conduct online A/B tests to demonstrate the effectiveness of our approach.

% TK

\subsection{Online A/B Tests}

The systematic scaling of our models yielded substantial cumulative gains across all key 
performance indicators. As shown in Table~\ref{tab:onlineexp}, a combined +2.1\% increase in 
offline mAP was observed alongside a +2.6\% lift in search conversion 
rate, as measured via online A/B testing. To remain latency-neutral, 
we employed a phased launch strategy, scaling model complexity to 8x %8\times$ 
FLOPs in tandem with dedicated inference optimization. Notably, while a cold-start 
approach would have incurred a +264\% training time overhead, our 
streamlined warmstart strategy reduced this to just +36\%, 
unlocking higher iteration frequency and velocity 
of model updates.

\begin{table}[!t]
    \centering
    %\resizebox{\linewidth}{!}{
    \begin{tabular}{|l|c|c|c|}
        \hline
        \multirow{2}{*}{\bf Experiment} & \bf $\Delta$ Offline & \bf $\Delta$ Online Search & \bf Inference \\
        & \bf mAP & \bf Conversion Rate &\bf FLOPs \\ 
        \hline
        Embedding + Backbone Scaling & 1.5\% & 1.9\% & \bf 8x\\
        Data Scaling & 0.6\% & 0.7\% & \bf 1x \\
        \hline
        \hline
        \textbf{Overall Impact} & \textbf{+2.1\%} & \textbf{+2.6\%} & \textbf{8x} \\ \hline
    \end{tabular}
    %}
    \caption{Gains of offline mAP and online search conversion rate from launches of backbone and data scaling.
    %Each model builds cumulatively on the previous version, demonstrating the compounded impact of systematic scaling. 
    Inference FLOPs are reported relative to the pre-scaling baseline.}
    \label{tab:onlineexp}
    %\vspace{-24pt}
\end{table}

%% file: tables/backbone-scaling-exp.tex
\begin{table*}[!t]
    \centering
    \resizebox{\linewidth}{!}{
        \begin{tabular}{c|c|c|cccc|cccc|cccc} \hline
        \multirow{2}{*}{Backbone} & Scaling & Factor Values & \multicolumn{4}{c|}{$\Delta$mAP} & \multicolumn{4}{c|}{$\Delta$Training Throughput} & \multicolumn{4}{c}{Inference FLOPs ($\times 10^7$)} \\ \cline{4-15}
        & Factor & V1, V2, V3, V4 & V1 & V2 & V3 & V4 & V1 & V2 & V3 & V4 & V1 & V2 & V3 & V4 \\ \hline
        
        \multirow{2}{*}{DCNv2} 
        & Cross Width & \textbf{1k}, 2k, 4k, 8k & +0.00\% & +0.12\% & +0.18\% & +0.18\% & +0.00\% & -1.85\% & -13.89\% & -39.63\% & 4.17 & 6.37 & 12.03 & 28.37\\ 
        & Deep Width & \textbf{1k}, 2k, 4k, 6k & +0.00\% & -0.01\% & +0.10\% & +0.15\% & +0.00\% & -7.41\% & -16.85\% & -38.33\% & 4.17 & 7.63 & 18.32 & 34.04 \\ \hline

        \multirow{4}{*}{MaskNet} 
        % & Cross Width & 1k, \textbf{2k}, 3k, 4k & +0.40\% & +0.52\% & +0.61\% & +0.64\% & +3.70\% & -0.93\% & -22.96\% & -27.13\% & 4.70 & 10.89 & 20.42 & 33.31 \\ 
        & Cross Width & 512, 1k, \textbf{2k}, 4k & +0.23\% & +0.40\% & +0.52\% & +0.64\% & +0.93\% & +3.70\% & -0.93\% & -27.13\% & 2.87 & 4.70 & 10.89 & 33.31 \\ 
        & Deep Width & 512, \textbf{1k}, 2k, 4k & +0.55\% & +0.52\% & +0.53\% & +0.54\% & 0.00\% & -0.93\% & -14.63\% & -18.61\% & 9.62 & 10.89 & 13.41 & 18.45 \\ 
        & Parallel Blocks & 1, \textbf{2}, 4, 8 & +0.50\% & +0.52\% & +0.56\% & +0.60\% & +5.56\% & -0.93\% & -19.81\% & -31.30\% & 6.27 & 10.89 & 20.12 & 38.59\\
        & Sequential Blocks & 1, 2, 4, 8 & +0.51\% & +0.51\% & +0.53\% & +0.42\% & -7.59\% & +0.00\% & -13.43\% & -28.61\% & 6.27 & 9.00 & 14.45 & 25.36 \\
        \hline
        
        \multirow{5}{*}{Transformer} 
        & $d_{\text{model}}$ & \textbf{32}, 64, 128, 192 & -0.01\% & +0.21\% & +0.25\% & +0.26\% & -1.85\% &  +0.93\% & -13.89\% & -33.33\% & 0.87 & 1.85 & 5.18 & 10.19 \\ 
        & Sequence Length & \textbf{16}, 32, 64, 128 & -0.01\% & +0.18\% & +0.33\% & +0.46\% & -1.85\% & -10.65\% & -18.33\% & -52.13\% &  0.87  & 1.75 & 4.22 & 12.01 \\
        & \# of Layers & 1, \textbf{2}, 4, 8 & -0.03\% & -0.01\% & +0.01\% & -0.03\% & +14.81\% & -1.85\% & -10.83\% & -22.04\% & 
        0.78  & 0.87 & 1.04 & 1.39 \\ 
        & \# of Heads & 1, \textbf{2}, 4, 8 & -0.01\% & -0.01\% & -0.02\% & +0.01\% & +0.93\% & -1.85\% & -14.91\% & -10.00\% & 
        0.83 & 0.87 & 0.93 &1.07 \\ 
        & $\text{ffn}_{\text{dim}}$ & 128, \textbf{256}, 512, 1k & -0.02\% & -0.01\% & -0.03\% & -0.04\% & +5.56\% & -1.85\% & -12.22\% & +2.78\% & 0.81 & 0.87 & 0.97 & 1.18 \\ 
        \hline

        \multirow{4}{*}{RankMixer} 
        & $d_{\text{model}}$ & \textbf{64}, 128, 256, 384 & +0.17\% & +0.37\% & +0.50\% & +0.53\% & -0.93\% & -20.09\% & -33.24\% & -51.11\% & 1.78 & 4.40 & 13.43 & 27.49 \\ 
        & Sequence Length & 8, \textbf{16}, 32, 64 & -0.02\% & +0.17\% & +0.34\% & +0.32\% & 0.00\% & -0.93\% & -22.69\% & -41.76\% & 0.99 & 1.78 & 39.80 & 10.91 \\ 
        & \# of Layers & 1, \textbf{2}, 4, 8 & +0.15\% & +0.17\% & +0.20\% & +0.21\% & -3.70\% & -0.93\% & -14.26\% & -27.04\% & 1.67 & 1.78 & 1.99 & 2.43 \\ 
        & $\text{ffn}_{\text{dim}}$ & 128, \textbf{256}, 512, 1k & +0.19\% & +0.17\% & +0.16\% & +0.17\% &  -7.78\% &  -0.93\% &  -5.56\% & -2.78\% & 1.67 & 1.78 & 1.99 & 2.42 \\ \hline
        \end{tabular}
    }
    \caption{Relative changes in mAP and training throughput across scaling factors with inference FLOPs against DCNv2 baseline. Bold factors are the default values for each backbone (the default MaskNet is configured without sequential blocks).}
    \label{tab:backbone_scaling_exp}
    %\vspace{-12pt}
\end{table*}

%% file: tables/iid-embedding-table.tex
\begin{table}[!h]
\centering
\begin{tabular}{c|cccc}
\hline
 & \multicolumn{4}{c}{\bf Embedding Dimension} \\ \cline{2-5}
\textbf{Vocab Size} & \textbf{16} & \textbf{32} & \textbf{64} & \textbf{128}  \\ \hline
1x                  & +0.00\%          & $+0.05\%$    & $+0.10\%$    & $+0.12\%$         \\
10x                 & $+0.02\%$    & $+0.09\%$    & $+0.13\%$    & $+0.16\%$     \\ \hline
\end{tabular}
\caption{Impact of scaling a MaskNet model on mAP from a baseline 1x vocab size and 16 embedding dimension.}
\label{tab:iid_embedding_vocab_size}
%\vspace{-20pt}
\end{table}

%% file: tables/missing-features-data-scaling.tex
\begin{table}[!t]
\centering
%\resizebox{\linewidth}{!}{

\if 0
\begin{tabular}{lcc}
%\toprule
\hline
\textbf{Setting} & \textbf{Training Days} & \textbf{mAP Gain} \\
%\midrule
\hline
\multirow{3}{*}{Including Missing Features}
 & 300d & 1.00\% \\
 & 400d & 1.02\% \\
 & 500d & - \\
\midrule
\multirow{3}{*}{Not Including Missing Features}
 & 300d & 0.97\% \\
 & 400d & 1.02\% \\
 & 500d & - \\
 \hline
%\bottomrule
\end{tabular}
\fi 
\begin{tabular}{c|cc} \hline
\multirow{2}{*}{\bf Training Days} & \multicolumn{2}{c}{\bf Coverage of Personalization Embeddings} \\ \cline{2-3}
    &  \bf Full Coverage & \bf Recent 300 days only \\ \hline
    300 & 1.00\% & 0.97\% \\
    400 & 1.02\% & 1.02\% \\ 
    500 & 1.05\%  & 1.03\%     \\
    \hline
\end{tabular}
%}

\caption{Summary of mAP gains for a MaskNet model with different sizes of training data and different coverage of personalization embeddings, which are  constructed by only historical item engagement without any personally identifiable information.}
\label{tab:missing_features_data_scaling}

%\vspace{-24pt}
\end{table}

%% file: figures/data-scaling-line-chart.tex
\begin{tikzpicture}
    \begin{axis}[
        xlabel={Duration of Training Data (days)},
        ylabel={$\Delta$mAP (\%)},
        xmin=30, xmax=310, % Set x-axis limits
        ymin=-0.1, ymax=1.75, % Set y-axis limits to accommodate all data
        xtick={35, 70, 140, 300}, % Define ticks on the x-axis
        ytick={0, 0.5, 1.0, 1.5, 2.0}, % Define ticks on the y-axis
        legend pos=south east, % Position the legend inside the plot area
        legend style={legend cell align=left},
        grid=major, % Add a major grid
        grid style={dashed, gray!30},
        width=8cm, % Set the width of the plot
        height=7cm, % Set the height of the plot
        ]
\if 0
        % Plot 1: DCNv2 with DNN Width 768
        \addplot[
            color=blue,
            mark=square,
            ]
            coordinates {
                (30, 0.00)
                (60, 0.79)
                (120, 1.42)
            };
        \addlegendentry{DCNv2 with DNN Width 768}

        % Plot 2: DCNv2 with DNN Width 6144
        \addplot[
            color=red,
            mark=triangle,
            ]
            coordinates {
                (30, 0.31)
                (60, 1.10)
                (120, 1.73)
            };
        \addlegendentry{DCNv2 with DNN Width 6144}
\fi
        % Plot 3: RankMixer with 64 d_model
        \addplot[
            color=green!60!black,
            mark=o,
            ]
            coordinates {
                (35, 0)
                (70, 0.40)
                (140, 0.67)
                (300, 0.72)
            };
        \addlegendentry{Transformer (Seq Len 16)}

        % Plot 4: RankMixer with 256 d_model
        \addplot[
            color=orange,
            mark=diamond,
            ]
            coordinates {
                (35, 0.37)
                (70, 0.87)
                (140, 1.24)
                (300, 1.49)
            };
        \addlegendentry{Transformer (Seq Len 128)}

        % Plot 5: MaskNet with 1 mask layer
        \addplot[
            color=purple,
            mark=star,
            ]
            coordinates {
                (35, 0.21)
                (70, 0.65)
                (140, 0.98)
                (300, 1.11)
            };
        \addlegendentry{MaskNet (Cross Width 512)}

        % Plot 6: MaskNet with 8 mask layers
        \addplot[
            color=blue,
            mark=square,
            ]
            coordinates {
                (35, 0.52)
                (70, 1.06)
                (140, 1.44)
                (300, 1.60)
            };
        \addlegendentry{MaskNet (Cross Width 4096)}
    \end{axis}
\end{tikzpicture}

%% file: figures/emb_data_scaling_line_chart_v2.tex
\begin{tikzpicture}
    \begin{axis}[
        xlabel={Duration of Training Data (days)},
        ylabel={$\Delta$mAP (\%)},
        xmin=30, xmax=150,
        ymin=-0.1, ymax=1.2,
        xtick={35, 70, 140},
        ytick={0, 0.2, 0.4, 0.6, 0.8, 1.0, 1.2},
        legend pos=south east,
        legend style={legend cell align=left},
%%        legend style={legend cell align=left, nodes={scale=0.8, transform shape}}, % Slightly scaled down legend
        grid=major,
        grid style={dashed, gray!30},
        width=8cm,
        height=7cm,
        ]

        % Plot 1: Emb dim 16
        \addplot[
            color=blue,
            mark=square,
            ]
            coordinates {
                (35, 0.00)
                (70, 0.48)
                (140, 0.87)
            };
        \addlegendentry{Emb dim 16}

        % Plot 2: Emb dim 32 (New)
        \addplot[
            color=green!60!black,
            mark=o,
            ]
            coordinates {
                (35, 0.05)
                (70, 0.52)
                (140, 0.96)
            };
        \addlegendentry{Emb dim 32}

        % Plot 3: Emb dim 64 (New)
        \addplot[
            color=orange,
            mark=diamond,
            ]
            coordinates {
                (35, 0.10)
                (70, 0.60)
                (140, 1.00)
            };
        \addlegendentry{Emb dim 64}

        % Plot 4: Emb dim 128
        \addplot[
            color=red,
            mark=triangle,
            ]
            coordinates {
                (35, 0.12)
                (70, 0.65)
                (140, 1.00)
            };
        \addlegendentry{Emb dim 128}

        % Plot 5: Emb dim 256 (New)
        \addplot[
            color=purple,
            mark=star,
            ]
            coordinates {
                (35, 0.14)
                (70, 0.68)
                (140, 0.99)
            };
        \addlegendentry{Emb dim 256}

    \end{axis}
\end{tikzpicture}

%% file: figures/cpu-vs-gpu-tikz.tex
\begin{tikzpicture}
    \begin{axis}[
        xlabel={Batch Size},
        ylabel={P99 Latency (ms)},
        legend style={at={(0.5,-0.4)}, anchor=north, legend columns=-1}, % Position legend outside and below the plot
        xmin=0, xmax=600, % Set x-axis limits
        ymin=0, ymax=100, % Set y-axis limits to accommodate all data
        ytick={0,25,...,100}, % Set y-axis ticks every 25
        xtick={0,50,...,600}, % Set x-axis ticks every 50 for readability
        %xtick={30, 60, 120}, % Define ticks on the x-axis
        %ytick={0,}, % Define ticks on the y-axis
        grid=major, % Add a major grid for better readability
        width=12cm, % Make the plot responsive to the text width
        height=4cm,
        xmin=0, % Start x-axis from 0
    ]

    % Plot for CPU
    % This plot contains data points that are not present in the GPU series.
    \addplot[
        color=blue,
        %mark=square*,
        %mark options={fill=blue!50!white},
        very thick,
        solid,
    ] coordinates {
        (1, 17.95165625)
        (5, 15.71223977)
        (10, 16.65113967)
        (20, 31.0143537)
        (30, 53.58232916)
        (60, 87.53753579)
        (120, 174.5781086)
    };
    \addlegendentry{CPU}

    % Plot for GPU (baseline)
    % This plot starts from batch size 10 and has more data points at the end.
    \addplot[
        color=red,
        %mark=triangle*,
        %mark options={fill=red!50!white},
        ultra thick,
        dashed,
    ] coordinates {
        (10, 67.09902609)
        (20, 66.38854108)
        (30, 67.24196548)
        (60, 68.21752039)
        (120, 68.20848719)
        (180, 70.30659252)
        (240, 71.7310241)
        (300, 74.85763217)
        (400, 76.37750407)
        (500, 80.58134528)
        (600, 82.10018276)
    };
    \addlegendentry{GPU (unoptimized)}

    % Plot for GPU (optimized)
    \addplot[
        color=green!60!black,
        %mark=*,
        %mark options={fill=green!50!white},
        ultra thick,
        dashdotted,
    ] coordinates {
        (10, 13.90699119)
        (20, 14.71893193)
        (30, 14.38039318)
        (60, 15.1160786)
        (120, 16.89893781)
        (180, 18.69459375)
        (240, 21.99078971)
        (300, 23.15967491)
        (400, 26.61430816)
        (500, 29.26411495)
        (600, 32.47045003)
    };
    \addlegendentry{GPU (optimized)}

    \end{axis}
\end{tikzpicture}

%% file: tables/server-side-batching.tex
% \begin{table}[!t]
% \centering
% % The table now has 4 columns.
% \resizebox{\linewidth}{!}{
% \begin{tabular}{|l|c|c|c|}
% \hline
% \textbf{Batch timeout} & \textbf{\makecell{Peak QPS \\ at 50ms}} & \textbf{\makecell{Peak QPS \\ at 75ms}} & \textbf{\makecell{Peak QPS \\ at 100ms}} \\
% \hline
% \makecell{0ms \\ (wo/ dynamic batching)} & 4.2k & 4.6k & 4.8k \\
% \hline
% 5ms & 14.0k & 17.1k & 17.2k \\
% \hline
% 10ms & 13.6k & 20.1k & 19.8k \\
% \hline
% 15ms & 13.8k & 19.9k & 19.8k \\
% \hline
% 20ms & 7.0k & 14.0k & 19.4k \\
% \hline
% 25ms & 6.1k & 11.7k & 20.5k \\
% \hline
% 30ms & 5.6k & 9.7k & 19.8k \\
% \hline
% \end{tabular}
% }
% \caption{Impact of server-side batching timeout on Peak QPS under different P99 latency constraints. QPS: thousands of items processed per second.}
% \label{tab:server_side_batching}
% \end{table}

\begin{table}[!h]
\centering
% The table now has 3 columns.
%\resizebox{\linewidth}{!}{
\begin{tabular}{c|cc|cc}
\hline
\textbf{Batch timeout} & \multicolumn{2}{c|}{\bf Peak QPS at 50ms} & \multicolumn{2}{c}{\bf Peak QPS at 75ms} \\
%\textbf{\makecell{Peak QPS \\ at 50ms}} & & \textbf{\makecell{Peak QPS \\ at 75ms}} \\
\hline
%\makecell{0ms \\ (w/o dynamic batching)} & 4.2k & 4.6k \\
0ms (w/o dynamic batching) & 4.2k & & 4.6k &\\ \hline
%\hline
5ms & \textbf{14.0k} & \textbf{3.33x} & 17.1k & 3.72x\\
%\hline
10ms & 13.6k & 3.24x & \textbf{20.1k} & \textbf{4.37x}\\
%\hline
15ms & 13.8k & 3.29x  & 19.9k  & 4.33x\\
%\hline
20ms & 7.0k & 1.67x  & 14.0k  & 3.04x\\
%\hline
25ms & 6.1k & 1.45x  & 11.7k  & 2.54x\\
%\hline
30ms & 5.6k & 1.33x  & 9.7k  & 2.11x\\
\hline
\end{tabular}
%}
\caption{Impact of server-side batching timeout on Peak QPS under different P99 latency constraints. QPS denotes thousands of items processed per second.}
\label{tab:server_side_batching}
%\vspace{-24pt}

\end{table}

%% file: contents/section5_relatedwork.tex
%\subsection{CVR Models}
%\subsection{Conversion Rate Prediction}

\vspace{-3pt}
\section{Related Work}
\label{section:relatedwork}

\mysection{Conversion Rate Prediction}
The prediction of conversion rate (CVR) is one of the most critical components in large-scale online e-commerce systems, such as product search~\cite{zhou2019deep} and advertising~\cite{zhu2021open}.
Unlike early-stage logistic regression~\cite{mcmahan2013ad} and decision tree~\cite{he2014practical} based CVR models that rely on hand-crafted features and feature engineering, recent deep learning models are trending towards learning hidden representations from raw input.
In particular, these models can be decomposed into three components, including feature extraction, feature interaction, and representation transformation~\cite{zhai2024actions}.
In this paper, we focus on examining the scaling capability for the feature interaction component and shared our learning on how to efficiently scale CVR models.

Existing studies utilize a variety of model backbones to represent feature interactions. A line of research~\cite{cheng2016wide,zhou2019deep} focuses on capturing higher-order interactions through the use of explicit feature crossing architectures~\cite{guo2017deepfm}. Moreover, Deep \& Cross Network~\cite{wang2017deep} and DCNv2~\cite{wang2021dcnv2} use the cross network for higher-order interactions among features, while MaskNet~\cite{wang2021masknet} leverages the instance-guided mask to learn cross-feature representations efficiently.
The rise of Transformers and self-attention~\cite{vaswani2017attention} further stimulates several works to model more comprehensive interactions~\cite{zhu2025rankmixer,zhang2024wukong}.
Furthermore, making ensembles of heterogeneous model backbones is also a popular research field~\cite{zhang2022dhen}.
In our study, we incorporate a variety of backbones to explore best practices for efficient model scaling.

%CVR models, which perform conversion rate prediction related to targets, have been a critical component of large scale online e-commerce systems. Starting from logstic regression and tree based models in the early stage, which requires lots of efforts on feature engineering, deep learning based models such as Wide\&Deep from Google, DIN from TaoBao is trending in the industry. Modern CVR models typically consist of three main components: Feature Extraction, Feature Interaction and Transformations of Representation(refer HSTU). Feature Extraction mainly transform categorical features into vectors through learned low-dimensional embedding.

%Then DCN and DCNV2 gained wide popularity in the industry. Gating based networks such as MaskNet are claimed to have better prediction performance than DCN. Recently, model ensemble such as DHEN has received lots of attention.

%Compared to existing work, this paper focuses on examining the scaling capability of existing model architecture for CVR prediction task, and shared practical lessons on how to scale up CVR models efficiently.

\mysection{CVR Model Improvement Levers}
Modern Search CVR models employ several specialized levers to optimize quality. To deal with label sparsity, the models leverage Multi-Task Learning to mitigate label sparsity by incorporating auxiliary signals like clicks and add-to-carts~\cite{ma2018mmoe, tang2020ple}. To handle high-cardinality features, efficient embedding hashing techniques are explored~\cite{Shi2020, Kang2021}. Some works explore architectures that naturally fuse sequential user actions with metadata like price and category~\cite{huang2026hyformer, zhu2025rankmixer}. Finally, the integration of Learning-to-Rank  shifts the objective from pointwise classification to listwise preference optimization, ensuring more accurate result ordering~\cite{qin2021neural}. As detailed in Section~\ref{section:prelim}, our architecture incorporates refined versions of each of these components. While further optimization of individual modules remains a subject for future work, the paper focuses on the impact of model scaling.

\mysection{CVR Model Scaling}
%\noindent \textbf{CVR Model Scaling.}
To improve CVR models, scaling has emerged as a primary lever for quality gains. Current research generally follows two trajectories. The first focuses on input scaling, where models are designed to ingest raw customer engagement sequences (often exceeding $O(10^5)$ actions) directly from logs~\cite{zhai2024actions}. The second focuses on architectural scaling, seeking to establish predictable scaling laws for model backbones \cite{xu2025climber}. Similar to the foundational work in LLMs~\cite{kaplan2020scaling, hoffmann2022training}, these studies indicate that performance follows a power-law relationship relative to model parameters, dataset size, and compute. Building on these laws, industry practitioners aim to reach Pareto frontiers by optimizing model stacking and identifying architectures with superior scaling properties \cite{zhang2022dhen, zhuang2025practice, zhang2024wukong}. Despite the traction, empirical evidence remains inconsistent. Some studies suggest that returns diminish rapidly as models scale \cite{yan2025scaling}. We posit that this discrepancy arises from a lack of granularity; unlike LLMs, which have well-defined scaling dimensions, CVR scaling remains under-explored. In this study, we provide a nuanced analysis of scaling search CVR models across three critical dimensions: model backbones, embeddings, and training data.

%% file: contents/section6_conclusion.tex
\vspace{-6pt}
\section{Conclusion}
\label{section:conclusion}

In this paper, we provide an effective and general mechanism to scale modern search CVR prediction systems with an empirical study.
First, we identify the independence of scaling effects across variables so that we can accelerate exploration by individually examining each feasible dimension.
Second, the trends of data scaling can be consistent across different model backbones, enabling a more efficient trajectory for model scaling exploration.
Last but not least, we also present infrastructure optimization needed for online serving of scaled CVR models, as well as warmstart for training performance optimization.
Online A/B tests show that successful model deployments significantly enhance customer engagement business metrics, notably a 2.6\% gain in search conversion rate.

The insights of this work can be concluded as follows:
(1) Mileage may vary when it comes to scaling approaches. It may or may not work in particular cases, such as some factors of backbone scaling in our study. More exhaustive exploration is essential to identify the best scaling recipe.
We expedite the exploration by identifying promising candidates through individual model scaling and relying on an empirical observation that a model would consistently outperform another model if it performs better with smaller training data.
%(2) Composition of training data can matter when it comes to the conversion prediction task. Larger data volume can lead to improvements, and it is also important to balance sufficient recency of data with diversity of user behaviors across a longer time range.
(2) Although DHEN-based ensemble models consistently outperform their individual models in our study, optimized models with a single model backbone can still outperform ensemble models under the same FLOPs budget.
In terms of long-term benefits, we suggest focusing on scaling single model backbone over ensemble-based approach for both simplicity and ease of recurring model development.
%Ensembling heterogeneous models can result in improved model performance, but it is not always a free lunch. Some combinations could even deteriorate while the model parameter count increases.
%This highlights the need for future work to gain a deeper understanding as to why certain ensembles provide gains while others do not.
(3) To accelerate training updates, warmstart is an efficient approach to speed up the training efficiency and reduce resource consumption.
(4) A combination of inference optimization strategies, such as hardware selection, execution graph placement, and request batching, may be required in order to successfully meet online serving constraints of scaled-up CVR models.